\documentclass[prb,twocolumn,preprintnumbers,superscriptaddress,showpacs]{revtex4-1}
\usepackage{graphicx}
\usepackage{amsmath}
\usepackage{amssymb}
\usepackage{xcolor}

\newcommand{\tr}{\ensuremath{{\mathbf{tr}}}}

\renewcommand{\d}{\ensuremath{{\mathrm{d}}}}

\makeatletter
\newcommand*\wt[1]{\mathpalette\wthelper{#1}}
\newcommand*\wthelper[2]{%
        \hbox{\dimen@\accentfontxheight#1%
                \accentfontxheight#11.15\dimen@
                $\m@th#1\widetilde{#2}$%
                \accentfontxheight#1\dimen@
        }%
}
\newcommand*\accentfontxheight[1]{%
       \fontdimen5\ifx#1\displaystyle
                \textfont
        \else\ifx#1\textstyle
                \textfont
        \else\ifx#1\scriptstyle
                \scriptfont
        \else
                \scriptscriptfont
        \fi\fi\fi3
}
\makeatother

\makeatletter
\def\@fnsymbol#1{\ensuremath{\ifcase#1\or \dagger\or \ddagger\or
    \mathsection\or \mathparagraph\or \|\or **\or \dagger\dagger
    \or \ddagger\ddagger \else\@ctrerr\fi}}
\makeatother

\begin{document}

\title{Doubling Theorem and Boundary States of Five-Dimensional Weyl Semimetal}
\author{Jing-Yuan~Chen}
\affiliation{Stanford Institute for Theoretical Physics, Stanford University, Stanford, CA 94305}
\author{Biao~Lian}
\affiliation{Princeton Center for Theoretical Science, Princeton University, Princeton, New Jersey 08544}
\author{Shou-Cheng~Zhang}
\thanks{Deceased 1 December 2018.}
\affiliation{Stanford Institute for Theoretical Physics, Stanford University, Stanford, CA 94305}
\affiliation{Stanford Center for Topological Quantum Physics, Stanford University, Stanford, CA 94305}

\begin{abstract}
We study the generic band structures of the five-dimensional (5D) Weyl semimetal, in which the band degeneracies are 2D Weyl surfaces in the momentum space, and may have non-trivial linkings with each other if they carry nonzero second Chern numbers. We prove a number of theorems constraining the topological linking configurations of the Weyl surfaces, which can be viewed as a 5D generalization of the celebrated Doubling Theorem for 3D Weyl semimetal. As a direct physical consequence of these constraints, the 5D Weyl semimetal hosts a rich structure of topological boundary states. We show that on the 4D boundary of the 5D Weyl semimetal, there are 3D chiral Fermi hypersurfaces protected by bulk Weyl surfaces. On top of that, for bulk Weyl surfaces that are linked and carry nonzero second Chern numbers, the associated boundary 3D Fermi hypersurfaces will shrink to singularities at certain energies, which trace out a protected 1D Weyl nodal arc, in analogy to the Fermi arc on the 3D Weyl semimetal surface.
\end{abstract}

\maketitle

\section{Introduction}

Topological states of matter are known as quantum states protected by topological invariants. Depending on whether the bulk states are gapped or not, they can be divided into gapped topological states and gapless topological states. While gapped topological phases have been extensively studied \cite{Fu2007,Qi2008,Qi2011,Hasan2010}, the study of gapless topological states in various dimensions is still less understood and ongoing. In general, the spatial dimension and symmetries largely determines the classification of topological states \cite{Kitaev2009,Ryu2010}. In three dimensions (3D), there are two well-known classes of gapless topological states: one class is the 3D Weyl semimetal \cite{Balents2011,Wan:2011udc,Murakami2007,Huang2015,Weng2015,
Liu2014,Zyuzin2012,Liu2013,Son2013,Xu2015a,Xu2015b}, which contains Weyl nodes in the band structure and need no symmetry protection other than the lattice translational symmetry. Each Weyl node is a $2$-fold degenerate point protected by the first Chern number, and gives rise to topologically protected Fermi arcs on the boundaries \cite{Berry1984,Wan:2011udc}. The other class is the nodal line semimetal or superconductor, which contain 1D nodal lines in the Brillouin zone (BZ) protected by crystal symmetry or chiral symmetry. The nodal lines may also have linking invariants \cite{Zhao2013,Schnyder2015,Shunji2013,Yip2014,Chiu2014,sun2017double, chen2017topological, ezawa2017topological, yan2017nodal, bi2017nodal, chang2017topological}, which give rise to to linking related topological responses \cite{Lian:2016avn}.

Extending the scope into general spatial dimensions has been proven a valuable approach in understanding gapped topological states \cite{Zhang2001,Qi2008,Kitaev2009,Ryu2010}. This approach has also been employed in the study of gapless topological states.
In spatial dimensions higher than three, an intriguing gapless topological state is the 5D Weyl semimetal, which is simultaneously characterized by both the second Chern numbers and the 5D linking invariants \cite{Lian:2016tbn,Lian:2017thp}.
Unlike the Weyl nodes in 3D Weyl semimetal which are zero dimensional, the band degeneracy submanifolds in 5D Weyl semimetals become 2D closed surfaces in the momentum space, which are called the \emph{Weyl surfaces} (WSs). Since the WSs are extended objects, they admit non-trivial global configurations, and in particular, linking numbers among each other. Remarkably, it is shown that the sum of linking numbers of a WS is equal to the second Chern number defined via the band Berry curvature \cite{Lian:2016tbn}.
Like the bulk-boundary correspondence of usual topological states of matter,
the WS linking numbers in the bulk of 5D Weyl semimetal protect topological 1D \emph{Weyl arcs} on the boundary of the system, as can be seen in explicit lattice models \cite{Lian:2016tbn}. Moreover, the WS linking and the second Chern number are closely related to Yang monopole \cite{Yang1978,Zhang2001} in 5D when the system restores a $\mathsf{TP}$ symmetry \cite{Lian:2017thp}. Transition between gapped topological phases in 5D can be also understood by having the gapless 5D Weyl semimetal as intermediate stage \cite{Lian:2017thp}, in analogy to the topological phase transitions in 3D \cite{Murakami2007}.

One of the most well-known theorem on 3D Weyl nodes is the Doubling Theorem \cite{Nielsen:1980rz, Nielsen:1981xu}, which restricts the Weyl nodes in the BZ to appear in pairs of opposite chirality. This theorem has played important role in the historical development of lattice quantum chromodynamics and Weyl semimetal. It is natural to ask whether a similar ``doubling" constraint exists for the WSs in a 5D Weyl semimetal.
While simple models of 5D Weyl semimetals have been constructed \cite{Lian:2016tbn,Lian:2017thp}, the generic constraints of WSs in 5D have not been carefully studied yet. The aim of this paper is to answer this question in a generic way.
We prove that there are non-trivial topological constraints governing the global configuration of the WSs and their linking. As we shall show, some of the constraints in 5D appear similar to the Doubling Theorem in 3D, however, the origins of which involve some substantial differences. Based on these constraints, we develop the correspondence between the bulk band topology and the gapless boundary states in a generic manner (as opposed to going into specific models), much like how the Doubling Theorem underlies the bulk-boundary correspondence in 3D Weyl Semimetal.

We should note that extended band degeneracy submanifolds can also appear in other systems, most notably nodal line semimetals and superconductors in 3D, and the nodal lines may also have linking and other topological consequences.
However, in these systems the extendedness of the degeneracy is protected by discrete spacetime symmetries (space group symmetry, etc.), in contrast to 5D Weyl semimetal in which the extendedness of the WSs is robust without symmetry (as long as lattice momentum is well-defined). Therefore, in general, the nodal lines in 3D systems are not subjected to stringent topological constraints as the Doubling Theorem or those that we are going to prove. Notably, however, when a 3D nodal line system carries non-trivial $\mathbb{Z}_2$ monopole charge \cite{Ahn:2018qgz, Wu:2018fgt}, the nodal line configurations in 3D are subjected to similar constraints as our Weyl surfaces in 5D; we will discuss the connection as we proceed.

This paper is organized as follows. In Sec. \ref{Sec3D} we review the well-known Doubling Theorem of Weyl nodes in 3D. In Section \ref{sect_Review} we review the notion of the Weyl surfaces in 5D Weyl semimetal, and the mathematical description of their linking. From this discussion we raise a few questions, which lead to the topological constraints we present and prove in Section \ref{sect_Constraints}. Moreover, we will discuss how these constraints are related to the non-abelian Yang monopoles. In Section \ref{sect_Boundary} we consider a 5D Weyl semimetal with a 4D physical surface, which hosts rich surface states protected by the topological constraints in the 5D bulk. Finally, we conclude by a few further remarks.

\section{The 3D Doubling Theorem}\label{Sec3D}

We first recall the physics of 3D Weyl semimetals and the Doubling Theorem for the Weyl nodes. We consider a band theory single-electron Hamiltonian $H(k)$ with $N$ bands, labeled by momentum $k$ in the BZ of $D=3$ dimensions. We assume there is no symmetry other than lattice translational symmetry and time translational symmetry. The $N\times N$ Hamiltonian can be generically diagonalized as
\begin{align}
H^\alpha_{\ \beta}(k) = \sum_{n=1}^N u_n^\alpha(k) \ E_n(k) \ {u_n^\ast}_\beta(k)\ ,
\label{Hamiltonian_diag}
\end{align}
which we may write as matrix factorization $H=UEU^\dagger$ for short. Without loss of generality, one can sort the energy levels so that \cite{Nielsen:1980rz}
\begin{align}
E_n(k) \leq E_{n+1}(k)\ ,
\label{Energy_levels}
\end{align}
which holds for any momentum $k$. Generically, without additional symmetry, the equality $E_n(k)=E_{n+1}(k)$ takes place only on a $(D-3)$-dimensional sub-manifold in the BZ. This is because when considering the degeneracy between two adjacent bands, one may project the Hamiltonian onto those two bands and use the Pauli matrices basis
\begin{align}
H_{proj}(k)=h_0(k)+h_1(k) \sigma^1 + h_2(k) \sigma^2 +h_3(k) \sigma^3.
\end{align}
The two bands are degenerate if and only if $h_1=h_2=h_3=0$. This yields three conditions to be satisfied by the $D$ components of $k$. When $D=3$, the degeneracy takes places at points, which are known as \emph{Weyl nodes}. We shall denote the positions of the Weyl nodes between the $n$th and $(n+1)$th band by $k_{n+1/2}^i$ (the superscript labels each Weyl node and the subscript indicates between which two bands it lies). Each Weyl node is associated with a chirality $c_{n+1/2}^i=\pm 1$ (right- or left-handed), as determined by the Hamiltonian in the following way. Define the abelian Berry connection and Berry curvature of the $n$th band as (we use the notation of differential forms and matrix multiplication)
\begin{align}
A_n(k) &\equiv -i \, u_n^\dagger(k) \: \d u_n(k), \nonumber \\[.2cm]
F_n(k) &\equiv \d A_n(k) = -i \, \d u_n^\dagger(k) \: \d u_n(k).
\end{align}
A Weyl node can then be viewed as a ``monopole'' of Berry curvature:
\begin{align}
\d F_n(k) =& -\star\sum_i \frac{c_{n+1/2}^i}{2} \: 4\pi \delta^3(k-k_{n+ 1/2}^i) \nonumber \\[.2cm]
& + \star\sum_j \frac{c_{n-1/2}^j}{2} \: 4\pi \delta^3(k-k_{n-1/2}^j)
\label{3D_monopole}
\end{align}
where now $i$ runs over the Weyl nodes between the $n$th and $(n+1)$th band, and $j$ runs over those between the $(n-1)$th and $n$th band; $\star$ denotes Hodge dual. To better understand this expression, we can draw a small sphere $S^i$ around a Weyl node $k_{n+1/2}^i$, then the \emph{first Chern number} is given by the chirality:
\begin{align}
(C_1)_n^{S^i} \equiv \oint_{S^i} \frac{F_n}{2\pi} = -c_{n+1/2}^i
\end{align}
and moreover, $(C_1)_{n+1}^{S^i} = -(C_1)_n^{S^i}=c_{n+1/2}^i$. The ``monopole'' property \eqref{3D_monopole} can be shown by linearizing $H_{proj}$ in $k-k_{n+1/2}^i$ near $k_{n+1/2}^i$, and performing the explicit calculation.

The famous Doubling Theorem \cite{Nielsen:1980rz, Nielsen:1981xu} asserts that, the total chirality of the Weyl nodes between the $n$th and the $(n+1)$th band satisfies
\begin{align}
\sum_i c_{n+1/2}^i = 0\ .
\label{3D_Doubling_Thm}
\end{align}
Namely, Weyl nodes have to appear in pairs of opposite chiralities. The proof is rather straightforward. Consider small spheres $S^i$ enclosing each $k_{n+1/2}^i$ and also $S^j$ enclosing each $k_{n-1/2}^j$. By Stoke's Theorem and Eq.~\eqref{3D_monopole}, we have
\begin{align}
\sum_{i, j} \oint_{S^{i, j}} \frac{F_n}{2\pi} = -\sum_i c_{n+1/2}^i + \sum_j c_{n-1/2}^j\ .
\label{3D_small_spheres}
\end{align}
On the other hand, the BZ is a closed manifold, so we can equally well view the ``outside'' of all the spheres as the ``inside'' (up to a minus sign from reversing orientation). Then the integral of $F_n$ is identically $0$, since now the ``inside'', used to be the ``outside'', contains no Weyl nodes. This indicates $\sum_i c_{n+1/2}^i = \sum_j c_{n-1/2}^j$. Besides, the lowest band $n=1$ cannot have any Weyl nodes connected from below, as there are no lower bands, thus we have $\sum_j c_{1/2}^j=0$. By iteration, we then arrive at the Doubling Theorem in Eq.~\eqref{3D_Doubling_Thm}.

The Doubling Theorem leads to important physical consequences in 3D Weyl semimetal, most remarkably the \emph{Fermi arc} on the surface of the semimetal \cite{Wan:2011udc}. More precisely, the 2D spatial boundary of the 3D Weyl semimetal has an associated 2D momentum space. The Weyl nodes in the 3D momentum space can be projected on the this 2D momentum space. It has been shown that the projected Weyl nodes must pair up with opposite chiralities, such that between each pair there connects a 1D Fermi arc, perpendicular to which a chiral boundary mode flows (since the mode is chiral, the 1D Fermi surface is an arc instead of a closed loop). This chiral surface mode is protected by the $C_1$ of the Weyl nodes at the ends of the Fermi arc. Therefore, the Doubling Theorem in the bulk give rise to well-defined topologically protected surface states on the surface of the system.

\section{Weyl Surfaces and Linking in 5D}
\label{sect_Review}

In this section we review the notion of Weyl surfaces (WSs) in a $D=5$ dimensional Weyl semimetal, and the mathematical relation between their linking number and the second Chern number \cite{Lian:2016tbn}. Then we motivate the question of topological constraints for WS linking, which we shall prove in the next section.

The Hamiltonian and its band energies again take the form of Eqs.~\eqref{Hamiltonian_diag} and \eqref{Energy_levels}, except that the momentum $k$ lives in $D=5$ dimensions. As we mentioned below Eq.~\eqref{Energy_levels}, the two-band degeneracy $E_n(k)=E_{n+1}(k)$ will generically take place on a $(D-3)$-dimensional submanifold in the $D$ dimensional BZ. By a similar argument, a three-band degeneracy $E_{n-1}(k)=E_n(k)=E_{n+1}(k)$ will generically take place on a $(D-8)$-dimensional sub-manifold. Therefore, in $D=5$ under consideration, we only have $2$D submanifolds of two-band degeneracies, which is called the WSs. To be specific, we denote the WS between the $n$th band and $(n+1)$th band as $W_{n+1/2}$, which is generically a 2D closed manifold. It may consist of multiple disjoint connected components, and we can denote each connected component by $W_{n+1/2}^i$, with $\cup_i W_{n+1/2}^i = W_{n+1/2}$. Similar to the Weyl node in 3D, the WS in $5$D can also be viewed as a ``monopole'' of Berry connection in 5D \cite{Lian:2016tbn}, satisfying an equation analogous to Eq.~\eqref{3D_monopole}:
\begin{align}
\d F_n(k) = 2\pi \left(\int_{k' \in W_{n-1/2}} - \int_{k' \in W_{n+1/2}} \right) \star\delta^5(k-k').
\label{5D_monopole}
\end{align}
The left-hand-side is a $3$-form, so is the right-hand-side, because $\star\delta^5(k-k')$ is a $5$-form and $\int_{k'\in W_{n\pm 1/2}}$ are $2$D integrals. This fixes the orientation on each connected WS $W_{n+1/2}^i$, similar to that Eq.~\eqref{3D_monopole} fixes the chirality $c_{n+1/2}^i$ for each 3D Weyl node.

Now we consider the linking between a WS $W_{n+1/2}^i$ and another WS $W_{n-1/2}^j$. In general, two sub-manifolds can link with each other if their sum of dimensions is equal to the total spatial dimension minus one (in this case $2+2=5-1$), so the WSs can indeed be linked together. An easy way to characterize their linking is the following: let $W_{n+1/2}^i = \partial \Sigma_{n+1/2}^i$ (the existence of such a 3D manifold $\Sigma_{n+1/2}^i$ will be proven in the next section), then $W_{n+1/2}^i$ and $W_{n-1/2}^j$ will have linking number one if there is an intersection point of $\Sigma_{n+1/2}^i$ and $W_{n-1/2}^j$. This is a straightforward generalization of the picture of linked loops in 3D.

\begin{figure}
\includegraphics[width=.25\textwidth]{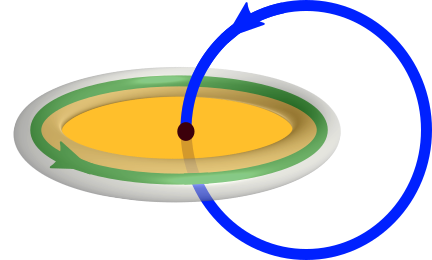}
\caption{Illustration of the computation relating the second Chern number to the linking number. The green loop represents the 2D WS component $W_{n+1/2}^i$, bounding the yellow region which represents the 3D $\Sigma_{n+1/2}^i$. The blue loop represents the 2D WS component $W_{n-1/2}^j$. The grey torus represents the 4D ``narrow tube'' $\partial V$, whose ``inside'' $V$ contains the green loop. The linking number is witnessed by the black dot, the intersection point of $\Sigma_{n+1/2}^i$ and $W_{n-1/2}^j$.}
\label{fig_linking_computation}
\end{figure}

It has been shown that the linking number between $W_{n+1/2}^i$ and $W_{n-1/2}^j$ is related to the \emph{second Chern number} of the Berry curvature in the $n$-th band \cite{Lian:2016tbn}. Here we show this using the above geometric picture of linking. Let's focus on the two WS connected components $W_{n+1/2}^i$ and $W_{n-1/2}^j$ which link with each other. Let $\partial V$ be a 4D ``narrow tube'' (which is topologically $W_{n+1/2}^i \times S^2$) enclosing $W_{n+1/2}^i$ which is the boundary of a 5D region $V$. We can choose $V$ narrow enough so that $V$ does not touch any other WSs in addition to $W_{n+1/2}^i$. The second Chern number on $\partial V$, viewed in the Berry curvature of the $n$-th band, is defined as
\begin{align}
(C_2)_n^{\partial V} \equiv \oint_{\partial V} \frac{1}{2} \left(\frac{F_n}{2\pi}\right)^2.
\end{align}
Using Stoke's Theorem and \eqref{5D_monopole}, we have
\begin{align}
(C_2)_n^{\partial V} = \int_V \frac{\d F_n}{2\pi} \frac{F_n}{2\pi}= -\oint_{W_{n+1/2}^i} \frac{F_n}{2\pi}\ ,
\label{C2_intermediate}
\end{align}
which becomes a 2D integral on WS $W_{n+1/2}^i$. Now we have a problem, since $F_n$ is singular on WS $W_{n+1/2}^i$. This is resolved as follows: since ``monopole'' is essentially a solution to the Poisson's equation, one can separate the Berry curvature as $F_n = F_n^{(1)}+F_n^{(2)}$, such that $F_n^{(2)}$ is non-singular on $W_{n+1/2}^i$, while $F_n^{(1)}$ is singular on $W_{n+1/2}^i$ but non-singular on $W_{n-1/2}^j$. One may view $F_n^{(1)}$ and $F_n^{(2)}$ as produced by ``monopoles'' $W_{n-1/2}^j$ and $W_{n+1/2}^i$, respectively. Then we should understand the last $F_n$ in Eq.~\eqref{C2_intermediate} as the non-singular part $F_n^{(2)}$ (in fact, the integral of the singular $F_n^{(1)}$ part gives self-linking number of $W_{n+1/2}^i$, which is identically zero in 5D. We will come back to this in the below) \cite{Lian:2016tbn}. Thus, using the Stoke's Theorem and Eq.~\eqref{5D_monopole} again we find
\begin{align}
(C_2)_n^{\partial V} &= -\int_{\Sigma_{n+1/2}^i} \frac{\d F_n^{(2)}}{2\pi} \nonumber \\[.2cm]
&= -\int_{k\in\Sigma_{n+1/2}^i} \int_{k' \in W_{n-1/2}^j} \star \delta^5(k-k') \nonumber \\[.2cm]
&\equiv -L_{W_{n+1/2}^i, W_{n-1/2}^j}
\label{C2_L}
\end{align}
where $L_{W_{n+1/2}^i, W_{n-1/2}^j}$ counts the (signed) number of intersection points of $\Sigma_{n+1/2}^i$ and $W_{n-1/2}^j$, and is thus the linking number between $W_{n+1/2}^i$ and $W_{n-1/2}^j$. This process of computation is illustrated in Fig.~\ref{fig_linking_computation}, where the linking number takes values $\pm 1$ depending on the orientation of the link.

The derivation in Eq.~\eqref{C2_L} motivates us to define two first Chern numbers associated with $W_{n+1/2}^i$ in the $n$-th band, the singular first Chern number $(C_1^s)_n$ and the regular first Chern number $(C_1^r)_n$:
\begin{equation}\label{C1sr}
(C_1^s)_n=\oint_{S^2}\frac{F_n^{(1)}}{2\pi}\ ,\quad (C_1^r)_n=\oint_{W_{n+1/2}^i}\frac{F_n^{(2)}}{2\pi}\ ,
\end{equation}
where $S^2$ in the definition of $(C_1^s)_n$ is an infinitesimal 2D sphere in the 3 co-dimensions of the WS $W_{n+1/2}^i$ which links with $W_{n+1/2}^i$. In general, $(C_1^s)_n$ is always $\pm1$, which also depend on the orientation of $S^2$. The second Chern number is then given by $(C_2)_n^{\partial V}=-(C_1^s)_n(C_1^r)_n$ (recall that topologically $\partial V \cong W_{n+1/2}^i \times S^2$). Similarly, one can define such two Chern numbers for $W_{n+1/2}^i$ in the $(n+1)$-th band, for which we choose the orientation of $S^2$ so that $(C_1^s)_{n+1}=(C_1^s)_{n}$, and we also have $(C_2)_{n+1}^{\partial V}=-(C_1^s)_{n+1}(C_1^r)_{n+1}$. Note that $(C_1^s)_n=\pm1$ simply implies the fact that a WS can be viewed as a Weyl node in the 3 co-dimensions orthogonal to the WS. As we shall prove in the next section, we always have $(C_2)_{n+1}^{\partial V}=-(C_2)_{n}^{\partial V}$, therefore under the above convention we have $(C_1^r)_{n+1}=-(C_1^r)_{n}$. The concept of these two first Chern numbers will appear in the surface state physics in Sec. \ref{sect_Boundary}.

The computation above shows a nice relation between the algebraic characterization $C_2$ and the geometric picture of linking. But it also raises several questions:
\begin{enumerate}
\item
We assumed every WS $W_{n+1/2}^i$ is a boundary $\partial \Sigma_{n+1/2}^i$ of some 3D manifold $\Sigma_{n+1/2}^i$. We need to prove such $\Sigma_{n+1/2}^i$ does always exist, so that the notion of ``linking'' and the computation above make sense.

\item
The linking we considered involve one WS component between the $(n-1)$th and $n$th band, and the other WS component between the $n$th and $(n+1)$th band. In principle, cannot one also consider both of them between the $n$th and $(n+1)$th band? (Geometrically one can also think of linking between, say, some $W_{n+1/2}^i$ and some $W_{n-3/2}^j$. But such linking is not detected by any Berry curvature and can be adiabatically pass through each other to unlink. So by ``linking'' we will not refer to such trivial possibilities.)

\item
There is an important distinction between 2D objects linked in 5D versus our familiar 1D objects linked in 3D. If one goes through the ``counting intersection points'' definition of $L$ carefully, one finds that, for 1D loops linked in 3D, $L_{C, C'}=L_{C', C}$, but for 2D surfaces linked in 5D, $L_{W, W'} = -L_{W', W}$ (such alternating sign repeats in higher odd dimensions mod 4D). \cite{Lian:2016tbn} This distinction has some important consequences in our problem:
\begin{enumerate}
\item In 3D (mod 4D), when a loop is properly regularized, it has a notion of \emph{self-linking}, which plays an important role in e.g. relativistic flux attachment. \cite{Polyakov:1988md} But in 5D (mod 4D), due to the negative sign above, the self-linking must vanish for self-consistency. \cite{Tze:1988ty} This justifies that in \eqref{C2_intermediate} we can regard the last $F_n$ as $F_n^{(2)}$ and drop $F_n^{(1)}$, instead of managing to regularize the $F_n^{(1)}$ contribution.

\item Suppose one generalizes the exercise \eqref{3D_small_spheres} from 3D to 5D -- that is, enclose each $W_{n+1/2}^i$ and each $W_{n-1/2}^j$ with a narrow tube, compute the $(C_2)_n$ over all these tubes, and then regard the ``outside'' of the tubes as the ``inside'', in order to derive any topological constraint about the links in the BZ. This exercise leads to that the total $(C_2)_n$ must be $0$, similar to \eqref{3D_Doubling_Thm}. While \eqref{3D_Doubling_Thm} implies Weyl nodes must appear in pairs of opposite chiralities, the vanishing of the total $(C_2)_n$ in the present case provides no topological constraint on the links -- the total $(C_2)_n$ being $0$ is always \emph{trivially} satisfied simply because $L_{W, W'} = -L_{W', W}$. Does this mean there is no topological constraint on the linking configurations in 5D Weyl semimetal?

\end{enumerate}

\end{enumerate}
These are the questions we will address next. Moreover, we will see the answers to these questions are directly related to the structure of the gapless surface states of a 5D Weyl semimetal, similar to that the Fermi arcs are related to the Doubling Theorem in 3D.

\section{Topological Constraints on Weyl Surfaces}
\label{sect_Constraints}

\subsection{Topological Constraints}
\begin{figure}
\includegraphics[width=.3\textwidth]{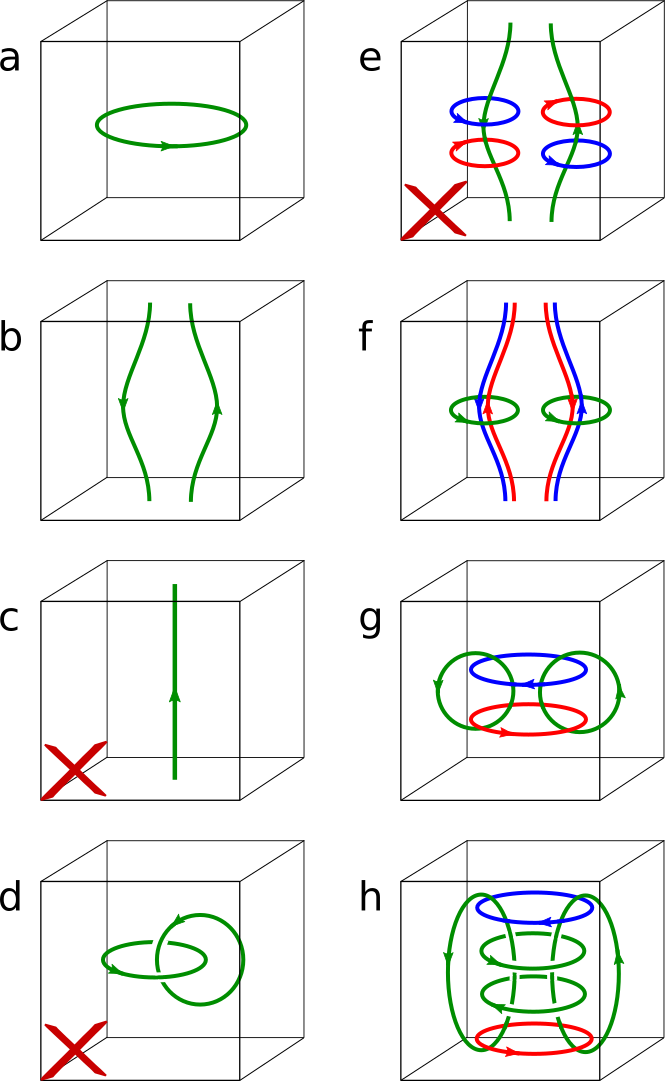}
\caption{Some examples of WS configuration satisfying / violating the topological constraints. We picture the 5D BZ by a 3D (periodic) cube, and 2D WS by 1D (oriented) loops. For simplicity we consider a 4-band system, with $W_{3/2}$ in blue, $W_{5/2}$ in green and $W_{7/2}$ in red. Configurations (a)(b)(f)(g)(h) are allowed, while (c)(d)(e) are not allowed for violating \eqref{5D_WS_boundary} \eqref{5D_WS_same} \eqref{5D_WS_upper_lower} respectively. The explicit lattice model constructed in Ref.~\cite{Lian:2016tbn} has a WS configuration (f). (A Hopf link seen by the $n$th band, like that in Fig.~\ref{fig_linking_computation}, is only allowed if the number of bands $N>4$, with $2<n<N-1$ due to \eqref{5D_WS_upper_lower}. Moreover, the total number of Hopf links seen by the $n$th band must be even, due to \eqref{5D_WS_Doubling}.)}
\label{fig_linking_constraint_ex}
\end{figure}

We shall show the following topological constraints on the WS and their linking:
\begin{enumerate}
\item
For each band index $n$, the WS $W_{n+1/2}$ must be a boundary,
\begin{align}
W_{n+1/2}=\partial\Sigma_{n+1/2}\ .
\label{5D_WS_boundary}
\end{align}

\emph{Remark:} $\Sigma_{n+1/2}$ may involve several connected components $\Sigma_{n+1/2}^i$, and we will denote $W_{n+1/2}^i = \partial \Sigma_{n+1/2}^i$. Note that we have modified our notation slightly -- in the previous section $W_{n+1/2}^i$ refers to a connected component of $W_{n+1/2}$, but now it might involve several connected components such that they together form the boundary of a connected component $\Sigma_{n+1/2}^i$. (For instance, in case (b) of Fig.~\ref{fig_linking_constraint_ex}, the two connected components together form a boundary, but not individually.)

Notice that Eq.~\eqref{5D_WS_boundary} resembles the 3D Doubling Theorem (a pair of Weyl nodes of opposite chiralities can be viewed as the ends of an arc), with Weyl nodes replaced by WS.

\item
For each $W_{n+1/2}^i$,
\begin{align}
L_{W_{n+1/2}^i, W_{n+1/2}} = 0,
\label{5D_WS_same}
\end{align}
i.e. its linking number with the other WS components between the same pair of bands is zero.

\item
For each $W_{n+1/2}^i$,
\begin{align}
L_{W_{n+1/2}^i, W_{n-1/2}} = -L_{W_{n+1/2}^i, W_{n+3/2}},
\label{5D_WS_upper_lower}
\end{align}
i.e. its linking numbers with the WS one band lower and the WS one band higher must be opposite.

\emph{Remark:} An important corollary is
\begin{align}
L_{W_{n+1/2}, W_{n-1/2}}=0,
\label{5D_WS_Doubling}
\end{align}
i.e. for each $n$, WS links must appear in pairs with opposite linking numbers. This is because \eqref{5D_WS_upper_lower} taking the union over $i$ leads to $L_{W_{n+1/2}, W_{n-1/2}} = -L_{W_{n+1/2}, W_{n+3/2}}$, then one can start with $n=1$ and iterate. Notice that  \eqref{5D_WS_Doubling} also resembles the 3D Doubling Theorem, but (in contrast to \eqref{5D_WS_boundary}) with Weyl nodes replaced by WS links.

\end{enumerate}
Some examples of allowed and not allowed WS configurations are illustrated in Fig.~\ref{fig_linking_constraint_ex}. The physical consequences of these constraints will be discussed in Sec. \ref{sect_Boundary}. In the rest of this section, we shall prove these constraints. (The connection between these constraints and ones in a special class of 3D nodal line systems \cite{Ahn:2018qgz} is discussed at the end of this Section.)

\subsection{Proof of the Topological Constraints}

The first constraint \eqref{5D_WS_boundary} can be proven using some formality. Clearly WS must be closed manifold. In the spirit of induction, suppose we have already shown $W_{n-1/2}$ is a boundary (the $n=1$ case is trivially true). Now suppose $W_{n+1/2}$ is not a boundary. Then there must exist some closed, non-exact differential 2-form $G$ such that $\oint_{W_{n+1/2}} G \neq 0$, for the following reason. The closed $W_{n+1/2}$ not being a boundary means it is a non-trivial element in the homology group $H_2(BZ)$. On the other hand, closed differential $m$-forms over the BZ are elements of the cohomology group $H^m(BZ, \mathbb{R})$. By the Universal Coefficient Theorem,
\begin{align}
H^m(BZ, \mathbb{R}) = \mathrm{Hom}(H_m(BZ), \mathbb{R})
\end{align}
where $\mathrm{Hom}$ is all homomorphisms. The right-hand-side for $m=2$ contains elements that map $W_{n+1/2}$ to non-zero real numbers, and thus guarantees the existence of $G$ with the said properties. With such $G$, we can use the familiar method \eqref{3D_small_spheres} from 3D. Consider narrow tube(s) $\partial V$ enclosing $W_{n+1/2}$, then by Stoke's Theorem, the closedness of $G$ and \eqref{5D_monopole}, we have
\begin{align}
\oint_{\partial V} \frac{F_n}{2\pi} \frac{G}{2\pi} = \int_V \frac{\d F_n}{2\pi} \frac{G}{2\pi} = -\oint_{W_{n+1/2}} \frac{G}{2\pi} \neq 0.
\end{align}
Now we regard the ``outside'' of $\partial V$ as the ``inside'' and apply the Stoke's Theorem again, we have
\begin{align}
-\oint_{\partial V} \frac{F_n}{2\pi} \frac{G}{2\pi} = \int_{BZ\backslash V} \frac{\d F_n}{2\pi} \frac{G}{2\pi} = \oint_{W_{n-1/2}} \frac{G}{2\pi} = 0
\end{align}
where the last equality is because $G$ is closed and, by induction assumption, $W_{n-1/2}$ is a boundary. Thus a contradiction arises. This shows $W_{n+1/2}$ must be a boundary.

The second constraint \eqref{5D_WS_same} is easily seen for a two-band system. Let $u_1$ be the first eigenvector. Then $F_1^2 = - (\d u_1^\dagger \d u_1)^2$. However, for a two-band system, $u_1$ and $u_1^\dagger$ together only has three real parameters (the phase choice can be fixed without affecting $F_1$), and therefore $F_1^2$, being a wedge product with four $\d$'s acting on three real parameters, vanishes. Thus, for a two-band system, the second Chern number around each $W_{3/2}^i$ -- whose integrand is proportional to $F_1^2$ -- must vanish, leading to \eqref{5D_WS_same}. For systems with more bands, the statement \eqref{5D_WS_same} only involves the $n$th and $(n+1)$th bands, so roughly speaking we can project the Hamiltonian as an effectively two-band system. However, in considering global topological effects such as linking, what we mean by ``effectively two-band'' needs to be carefully addressed. We leave this technical detail to Appendix A.

In the third constraint \eqref{5D_WS_upper_lower}, the left-hand-side is detected by $F_{n}$ and right-hand-side by $F_{n+1}$. We need to find a relation relating the two. Consider a narrow tube $\partial V$ enclosing a WS component $W_{n+1/2}^i$ in its ``inside''; we choose it to be so narrow that $V$ is disjoint from any other WS. We use the fact that the sum of Chern numbers over all bands vanishes:
\begin{align}
\sum_{n'=1}^N (C_2)_{n'}^{\partial V} = 0.
\end{align}
(This can be deduced by iterating \eqref{C2_NA_and_C2_A_general}, starting with $n=N$.) For $n'=n$, we have $(C_2)_{n}^{\partial V}=-L_{W^i_{n+1/2}, W_{n-1/2}}$. For $n'=n+1$, we have $(C_2)_{n+1}^{\partial V}=-L_{W^i_{n+1/2}, W_{n+3/2}}$. Other $(C_2)_{n'}^{\partial V}$ vanish because $W_{n+1/2}$ is not seen by other $F_{n'}$. This leads to \eqref{5D_WS_upper_lower}.

This derivation of \eqref{5D_WS_upper_lower} and hence \eqref{5D_WS_Doubling} appear very different from the familiar derivation of the 3D Doubling Theorem, despite that \eqref{5D_WS_Doubling} is a ``Doubling Theorem'' for WS links. In particular, the derivation above does not involve ``enclosing all the links and then viewing the `outside' as the `inside' ''. Is it possible to re-derive \eqref{5D_WS_Doubling} from such a perspective? Another question one may ask is, there is an interesting relation between WS link and Yang monopole \cite{Lian:2017thp}, but it seems not to manifest in steps above. In the below, we show the answers to these two questions are related. We first discuss the relation between WS link and Yang monopole through \emph{non-abelian Berry curvature}, and using this notion we provide an alternative direct derivation of \eqref{5D_WS_Doubling} that resembles the derivation of the 3D Doubling Theorem.

\subsection{Weyl Surface Link, Yang Monopole and Non-Abelian Berry Curvature}

The ``Doubling Theorem'' \eqref{5D_WS_Doubling} for WS links is an interesting result. In 3D nodal line materials in which linking also plays interesting role, \cite{Lian:2016avn, sun2017double, chen2017topological, ezawa2017topological, yan2017nodal, bi2017nodal, chang2017topological} there is no topological constraint enforcing the nodal line links to appear in pairs (unless more stringent constraint is applied \cite{Ahn:2018qgz, Wu:2018fgt} which we will return to in the end). How come in 5D the WS links must appear in pairs? This is because nodal line in 3D is protected by discrete symmetry, while WS in 5D is protected by topology. In the below we provide an alternative view towards the constraint \eqref{5D_WS_Doubling}. One can motivate \eqref{5D_WS_Doubling} by the relation between WS links and Yang monopoles \cite{Lian:2017thp}: a WS link can adiabatically arise from deforming a Yang monopole, and Yang monopoles themselves must appear in pairs, like Weyl nodes, so WS links must also appear in pairs. Now we proceed into the details.

While our previous proof of \eqref{5D_WS_Doubling} only employed abelian Berry curvature, in our alternative view to be introduced now we relate non-abelian Berry curvature to abelian ones. Let us introduce the notion of non-abelian Berry curvature. Recall the Hamiltonian is diagonalized as $H=UEU^\dagger$, and the eigenvalues in $E$ are ordered from low to high energies. The matrix $U$ has $N$ columns, being the eigenvectors $u_1, \cdots, u_N$. Let $U_{\leq n}$ be the $N\times n$ rectangular matrix consisting of the first $n$ columns of $U$, i.e. the columns of this rectangular matrix is $u_1, \cdots, u_n$. In other words, $U_{\leq n}$ is $U$ projected to the first $n$ bands. Note that $U_{\leq n} U_{\leq n}^\dagger$ is the projection matrix onto the first $n$ bands, and $U_{\leq n}^\dagger U_{\leq n}$ is the identity $\mathbf{1}_{n\times n}$ acting on the first $n$ bands. The non-abelian Berry connection
\begin{align}
A_{\leq n} \equiv -i \, U_{\leq n}^\dagger \: \d\, U_{\leq n}
\end{align}
is therefore an $n\times n$ matrix-valued connection $1$-form. The associated non-abelian Berry curvature
\begin{align}
F_{\leq n} &\equiv \d A_{\leq n} + iA_{\leq n}^2 \nonumber \\[.2cm]
&= -i \, \d U_{\leq n}^\dagger \left(1-U_{\leq n} U_{\leq n}^\dagger \right) \d U_{\leq n}
\end{align}
is an $n\times n$ matrix-valued curvature $2$-form. (One may wonder, in the absence of band degeneracy, the system does not have the $U(n)$ symmetry rotating among the first $n$ bands, why would such $U(n)$ Berry curvature still be useful to define. The idea is, this Berry curvature captures the separation of the original $N$-dimensional vector bundle over the BZ into two sub-bundles consisting of the first $n$ bands and the $N-n$ bands respectively. In doing so, we are viewing the first $n$ bands as a whole and not worrying about the further separation within them.) Note that $F_{\leq n}$ is only singular on the WS $W_{n+1/2}$ between the $n$th and the $(n+1)$th band, but does not see, say, $W_{n-1/2}$, in contrast to $F_n$, because $F_{\leq n}$ treats the first $n$ bands as a whole.

Having introduced the notion of non-abelian Berry curvature, now let's consider it in a 4-band system with $\mathsf{TP}$ symmetry, such as the model constructed in Ref.\cite{Lian:2017thp}. The first and second bands are completely degenerate, so are the third and fourth bands. Yang monopoles exist between the two pairs of degenerate bands in the 5D BZ. A Yang monopole is characterized by that the non-abelian second Chern number
\begin{align}
(C_2)_{\leq 2}^{S^4} \equiv \tr \int_{S^4} \frac{1}{2}  \left(\frac{F_{\leq 2}}{2\pi}\right)^2
\end{align}
around an $S^4$ enclosing the Yang monopole takes value $\pm 1$. The $\pm 1$ Yang monopoles appear in pairs in analogy to the 3D Doubling Theorem, due to an analogous proof. A $\mathsf{TP}$-breaking perturbation of this model does two things \cite{Lian:2017thp}: 1) it lifts the complete degeneracy between the first and second bands, leaving a WS degeneracy $W_{3/2}$ (and likewise for the third and fourth bands), and 2) it ``stretches'' each point-like Yang monopole into a small spherical WS component $W_{5/2}^i$  linked to $W_{3/2}$ and $W_{7/2}$. The resulting WS configuration is like example (f) of Fig.~\ref{fig_linking_constraint_ex}. If we turn on this $\mathsf{TP}$-breaking perturbation adiabatically,  $(C_2)_{\leq 2}^{S^4}$ should not change as long as the small $W_{5/2}^i$ sphere is still enclosed by the $S^4$.

\begin{figure}
\includegraphics[width=.46\textwidth]{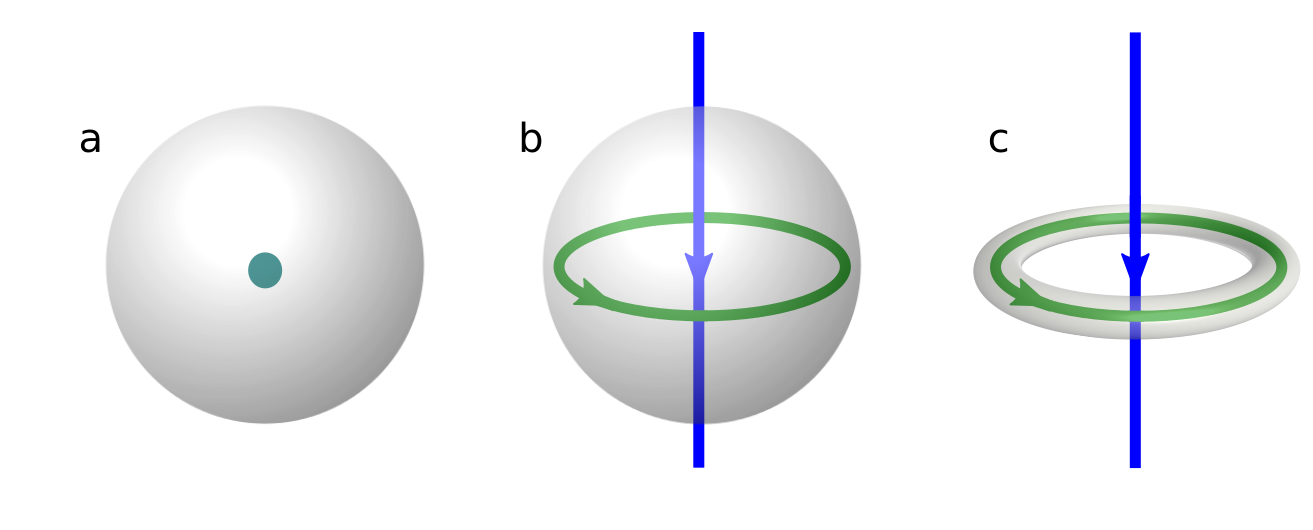}
\caption{The non-abelian second Chern numbers $(C_2)_{\leq n}^{S^4}$ in (a), $(C_2)_{\leq n}^{S^4}$ in (b) and $(C_2)_{\leq n}^{\partial V}$ in (c) are all equal to the abelian second Chern number $(C_2)_n^{\partial V}=1$ in (c). From (a) to (b) the Yang monopole is ``stretched out'' into a WS link due to $\mathsf{TP}$-breaking. From (b) to (c) the sphere $S^4$ enclosing $W_{n+1/2}^i$ is deformed to the narrow tube $\partial V$.}
\label{fig_Yang_WSlink}
\end{figure}

The example above motivates us to propose the following claim: $(C_2)_{\leq n}^{S^4}$ equals $\pm 1$ if the $S^4$ encloses a small WS component $W_{n+1/2}^i$ that links with $W_{n-1/2}$. Note that $W_{n-1/2}$ may intersect with the $S^4$, but this is no problem because $F_{\leq n}$, treating the first $n$ bands as a whole, does not see (i.e is non-singular on) $W_{n-1/2}$. But now that $F_{\leq n}$ only sees $W_{n+1/2}$, we can deform the $S^4$ to a narrow tube $\partial V$ enclosing $W_{n+1/2}^i$ (such that $V$ is disjoint from any other WS), without changing the value of $C_2$, i.e. $(C_2)_{\leq n}^{\partial V} = (C_2)_{\leq n}^{S^4}$. Thus, our claim becomes
\begin{align}
(C_2)_{\leq n}^{\partial V} \equiv \tr \int_{\partial V} \frac{1}{2}  \left(\frac{F_{\leq n}}{2\pi}\right)^2 = \pm 1.
\label{C2_NA}
\end{align}
This idea is illustrated in Fig.~\ref{fig_Yang_WSlink}. Moreover, with $S^4$ replaced by $\partial V$, we no longer need to require $W_{n+1/2}^i$ to be small so that it can be enclosed by an $S^4$; it can run across the BZ and still be enclosed by $\partial V$.

Let's prove the claim \eqref{C2_NA}. More particularly, we want to show the non-abelian second Chern number $(C_2)_{\leq n}^{\partial V}$ is in fact equal to the abelian one $(C_2)_n^{\partial V}$ that counts the linking number between $W_{n+1/2}^i$ and $W_{n-1/2}$. To show this equality, in Appendix B we show that, on a 4D closed manifold $\mathcal{M}$ that does not intersect $W_{n\pm 1/2}$, we have
\begin{align}
\tr \int_{\mathcal{M}} {F_{\leq n}}^2 = \int_{\mathcal{M}} {F_n}^2 + \tr \int_{\mathcal{M}} {F_{<n}}^2.
\label{C2_NA_and_C2_A_general}
\end{align}
(This is a special case of the Whitney sum formula for characteristic classes.) Taken $\mathcal{M} = \partial V$ where $V$ contains $W_{n+1/2}^i$ but disjoints from $W_{n-1/2}$, we get
\begin{align}
(C_2)_{\leq n}^{\partial V} = (C_2)_n^{\partial V}
\label{C2_NA_and_C2_A}
\end{align}
because the $F_{< n}$ does not see $W_{n+1/2}$. This is the general case of the result $(C_2)_{\leq 2}^{\partial V} = (C_2)_2^{\partial V}$ explicitly computed in the 4-band model in Ref.~\cite{Lian:2017thp}.

We can have a more intuitive understanding of the ``Doubling Theorem'' \eqref{5D_WS_Doubling} for WS links that resembles that of the 3D Doubling Theorem. Now let $\partial V$ be narrow tubes that enclose all components of $W_{n+1/2}$. By \eqref{C2_NA_and_C2_A}, we have $(C_2)_{\leq n}^{\partial V}=-L_{W_{n+1/2}, W_{n-1/2}}$. Next we view the ``outside'' of $\partial V$ as the ``inside'' and evaluate $(C_2)_{\leq n}^{\partial V}$ again; but there is no $W_{n+1/2}$ to be picked up, so the result must vanish (recall that by construction $F_{\leq n}$ only sees $W_{n+1/2}$, in contrast to $F_n$ which also sees $W_{n-1/2}$). This proves \eqref{5D_WS_Doubling}.

The relation to Yang monopole also gives an intuitive understanding of \eqref{5D_WS_upper_lower}. Upon $\mathsf{TP}$-breaking, a Yang monopole is ``stretched out'' into a component $W_{n+1/2}^i$ linked with $W_{n-1/2}$; but we can equally well replace $W_{n-1/2}$ with $W_{n+3/2}$, which resonates with \eqref{5D_WS_upper_lower}. This argument, however, cannot be taken as an alternative proof of \eqref{5D_WS_upper_lower} because \emph{a priori} we cannot assume each WS linking arises adiabatically from a Yang monopole.

Before we close, we would like to comment on an interesting connection between our constraints in 5D Weyl semimetal to those in certain 3D nodal line systems. It is shown \cite{Ahn:2018qgz, Wu:2018fgt} that in 3D nodal line systems protected by $\mathsf{TP}$ symmetry with $(\mathsf{TP})^2=+1$, one can define a $\mathbb{Z}_2$ monopole charge characterized by the second Stiefel-Whitney class of the band structure, and if the charge is non-trivial, the nodal lines must develop links, in a manner parallel to the constraints we proved in this section. Notably, one process to visualize \cite{Ahn:2018qgz} the relation between the $\mathbb{Z}_2$ monopole charge and the linking number is parallel to our visualization Fig. \ref{fig_Yang_WSlink}. Although the method employed there appears differently from ours, an intuitive explanation to the similarity between the constraints is that, the $(\mathsf{TP})^2=+1$ condition demands the Hamiltonian to be real valued, hence Refs.\cite{Ahn:2018qgz, Wu:2018fgt} and us are considering the same kind of constraint problems for real versus complex valued Hamiltonians, in 3D and 5D respectively.

\section{Topological Surface States}
\label{sect_Boundary}

The above topological constraints on WSs and their linkings lead to rich surface states on the 4D surface of a 5D Weyl semimetal, as we will derive in this section. In particular, we give a more intuitive and clearer understanding of the Weyl arc on the surface \cite{Lian:2016tbn,Lian:2017thp}, which is known to be protected by the second Chern number of WSs.

We first recall that the 3D Weyl semimetal hosts topologically protected Fermi arcs on its 2D surface, each of which connects the surface projection of two Weyl nodes with monopole charges (first Chern numbers) $+1$ and $-1$, respectively \cite{Wan:2011udc}. In other words, the constant electron energy contour of the surface states on a 2D surface of 3D Weyl semimetal is not a closed loop (the usual Fermi surface in 2D), but an open Fermi arc connecting two Weyl nodes with opposite monopole charges. Mathematically, one can show the first Chern number of a Weyl node requires its projection on the 2D surface to be connected with a Fermi arc \cite{Lian:2017thp}.

The 4D surface states of 5D Weyl semimetal exhibits a richer structure. There are two types of topological surface state features: 3D \emph{Fermi hypersurface} protected by the first Chern number (Eq.~\eqref{C1sr}) of the WS, and, on top of it, 1D \emph{Weyl arc} protected by the second Chern number (Eq.~\eqref{C2_intermediate}) of the WS. Now we explain them in details.

\begin{figure}
\includegraphics[width=.46\textwidth]{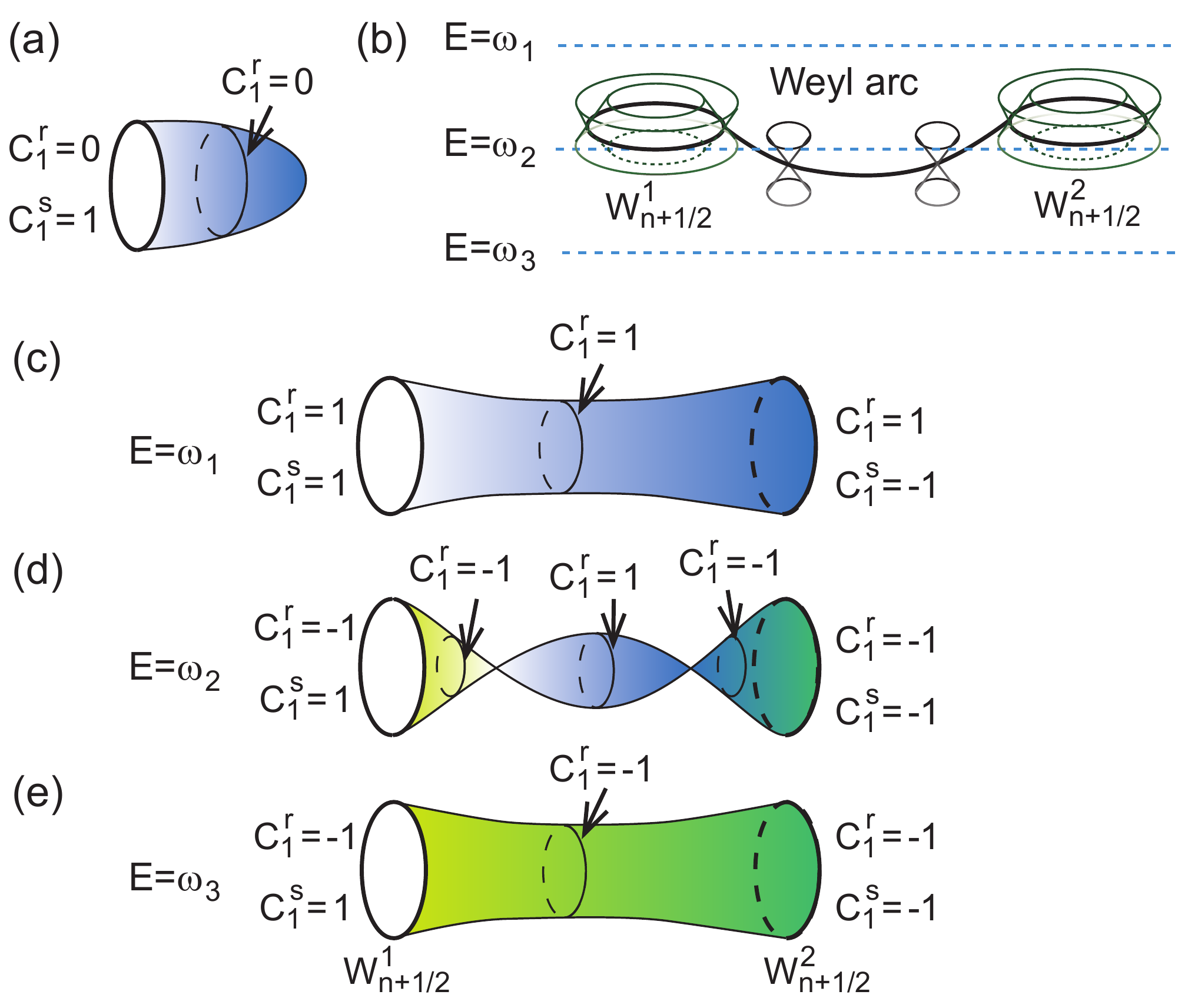}
\caption{Illustration of the topological surface states on the 4D surface of a 5D Weyl semimetal. (a) At a given fermi energy (near the WS), a WS $W_{n+1/2}^1$ with second Chern number $(C_2)_n^{\partial V}=0$ (which implies $C_1^s=1$ and $C_1^r=0$) protects a 3D Fermi hypersurface in the 4D surface momentum space, whose 2D boundary is the surface projection of the WS (the solid circle). (b) Illustration of the energy dispersion of the surface states topologically protected by two WSs $W_{n+1/2}^1$ and $W_{n+1/2}^2$ with second Chern numbers $(C_2)_n^{(1)}=-1$ and $(C_2)_n^{(2)}=1$, respectively. A Weyl arc arises in the surface states connecting the projection of the two WSs. (c)-(e) The topological surface states (3D Fermi hypersurfaces) at Fermi energies $E=\omega_1,\omega_2$ and $\omega_3$ defined in panel (b), whose boundaries (2D) are the surface projection of the two WSs (the left and right solid circles). The regular and singular first Chern numbers of the WSs are labeled by the side of each WS. At energy $E=\omega_2$, the constant energy plane intersects with the Weyl arc in panel (b), leading to two Weyl points on the Fermi hypersurface in panel (d).}
\label{fig_SS}
\end{figure}

First, consider a WS $W_{n+1/2}^i$ with its second Chern numbers $(C_2)_n^{\partial V}=-(C_2)_{n+1}^{\partial V}=0$. According to the constraints we revealed in last section, such a WS can exist alone, such as (a) and (b) in Fig.~\ref{fig_linking_constraint_ex}. It is, however, not completely topologically trivial, since it still carries a singular first Chern number $(C_1^s)_n=+1$ (recall \eqref{C1sr}). As we have explained in Sec. \ref{sect_Review}, the first Chern number $(C_1^s)_n$ simply implies each point of the WS behaves as a Weyl node in the 3 co-dimensions to the 2D WS. Therefore, while in 3D Weyl semimetal each 0D Weyl node when projected to the 2D physical surface serves as an end point of a 1D Fermi arc \cite{Wan:2011udc} due to the first Chern number's protection, for exactly the same mathematical reason, in 5D Weyl semimetal each 2D WS when projected to the 4D physical surface serves as the boundary of a 3D \emph{Fermi hypersurface}, which occurs at a given Fermi energy near the energy of the WS. Such a 3D Fermi hypersurface with chiral 2D boundary (the projected WS) is the topological surface state protected by $(C_1^s)_n$, and can only exist on the 4D surface of a 5D system, because otherwise any Fermi surface in an intrinsically 4D system is necessarily a 3D closed manifold. Fig.~\ref{fig_SS}(a) gives an illustration of the 3D Fermi hypersurface in the surface momentum space, where we have omitted one dimension of the 4D momentum space, so that the solid circle stands for the projected 2D WS, while the surface connected to it represents the 3D Fermi hypersurface. It is important to note that, in the present case, because $(C_2)_n^{\partial V}=-(C_2)_{n+1}^{\partial V}=0$, one has $(C_1^r)_n=(C_1^r)_{n+1}=0$  (recall \eqref{C1sr}), which is the Berry curvature integrated on the 2D WS. Accordingly, any 2D cross section on the 3D Fermi hypersurface which can continuously deform into (i.e., homotopic to) the 2D WS, e.g., the dashed circle in Fig.~\ref{fig_SS}(a), will have a first Chern number $C_1^r=0$, which is defined as the integration of Berry curvature of the 3D Fermi hypersurface states on the 2D cross section. Therefore, such a 2D cross section can shrink to zero, which means the 3D Fermi hypersurface can close by itself away from the WS, forming a half 3D sphere topologically as shown in Fig.~\ref{fig_SS}(a).

Then, we turn to WS $W_{n+1/2}^i$ with nonzero second Chern number $(C_2)_n^{\partial V}=-(C_2)_{n+1}^{\partial V}=1$ (according to the topological constraints we have proved, one always has $(C_2)_n^{\partial V}=-(C_2)_{n+1}^{\partial V}$), and show that nonzero second Chern numbers imply the existence of 1D \emph{Weyl arcs} \cite{Lian:2016tbn} on top of the 3D Fermi hypersurfaces.

Consider two WSs $W_{n+1/2}^1$ and $W_{n+1/2}^2$ as shown in Fig.~\ref{fig_SS}(b), which has second Chern numbers $(C_2)_n^{(1)}=-(C_2)_{n+1}^{(1)}=-1$ and $(C_2)_n^{(2)}=-(C_2)_{n+1}^{(2)}=1$, respectively. For simplicity, we assume they are the only WSs between the $n$-th band and the $(n+1)$-th band (as is allowed by the 5D doubling constraints). If we set the Fermi energy $E=\omega_1$ to be above the two WSs in the $(n+1)$-th band (Fig.~\ref{fig_SS}(b)), we will expect a 3D Fermi hypersurface on the 4D surface of the 5D semimetal as shown in Fig.~\ref{fig_SS}(c) (the left and right solid circles represent 2D WSs $W_{n+1/2}^1$ and $W_{n+1/2}^2$ projected on the 4D surface), which we shall explain below. Since the WS $W_{n+1/2}^1$ has singular first Chern number $(C_1^s)_{n+1}^{(1)}=1$, one would expect the projected $W_{n+1/2}^1$ to be boundary of a 3D Fermi hypersurface. However, when viewed in the $(n+1)$-th band, $(C_2)_{n+1}^{(1)}=1$ implies the WS also has a regular first Chern number $(C_1^r)_{n+1}^{(1)}=1$. Therefore, the surface states on a 2D cross section on the 3D Fermi hypersurface (the dashed circle in the middle) which is homotopic to $W_{n+1/2}^1$ will have the same regular first Chern number $C_1^r=1$. Since $C_1^r$ is nonzero, the 2D cross section cannot contract to zero by itself when moved continuously on the 3D Fermi hypersurface; instead it can only be continuously moved to another boundary of the Fermi hypersurface with regular first Chern number $C_1^r=1$, which in this case has to be the surface projection of the other WS $W_{n+1/2}^2$ (which under proper orientation choice has $(C_1^s)_{n+1}^{(2)}=-1$ and $(C_1^r)_{n+1}^{(2)}=1$). Therefore, the 3D Fermi hypersurface has to connect $W_{n+1/2}^1$ and $W_{n+1/2}^2$ as shown in Fig.~\ref{fig_SS}(c) due to nonzero second Chern number, which is clearly different from the case in Fig.~\ref{fig_SS}(a).

Now assume we lower the Fermi energy to $E=\omega_2$, which is slightly below the WSs and enters the $n$-th band. Upon entering into the $n$-th band, the second Chern number of a WS changes sign (relative to that viewed in the $(n+1)$-th band), so the regular first Chern number of $W_{n+1/2}^1$ flips sign to $(C_1^r)_{n}^{(1)}=-1$ (while $(C_1^s)_{n}^{(1)}=1$ remains unchanged) as shown in Fig.~\ref{fig_SS}(d), and similarly for $W_{n+1/2}^2$. Accordingly, $C_1^r$ of any 2D cross section on the 3D Fermi hypersurface that is homotopic to $W_{n+1/2}^1$ has to change sign, too. It may happen that certain 2D cross sections in the middle of the 3D Fermi hypersurface (the dashed circle in the middle in Fig.~\ref{fig_SS}(d)) still has $C_1^r=+1$ (unflipped). Then the cross sections with opposite $C_1^r=+1$ on the 3D Fermi hypersurface has to be connected via a Weyl point on the 3D Fermi hypersurface, as shown in Fig.~\ref{fig_SS}(d). Namely, two Weyl points will emerge from the two WSs and move towards each other on the 3D Fermi hypersurface as the Fermi energy is lowered. If one further lower the Fermi energy to $E=\omega_3$, the two Weyl points will annihilate with each other in the middle of the 3D Fermi hypersurface, and the $C_1^r$ of all 2D cross sections will be flipped, as shown in Fig.~\ref{fig_SS}(e). Therefore, because of the fact that the nonzero second Chern number $(C_2)_{n+1}^{(i)}=-(C_2)_{n}^{(i)}$ flips sign from the $n$-th band to the $(n+1)$-th band, the 3D Fermi hypersurface necessarily experience the arising (at two WSs) and annihilation of two Weyl points when the Fermi energy is changed from the $(n+1)$-th band to the $n$-th band.

If one plot the energy dispersion of the topological surface states on the 4D surface, one would expect to see the two Weyl points as a function of energy forming a 1D Weyl arc connecting the surface projection of the two WSs $W_{n+1/2}^1$ and $W_{n+1/2}^2$, as shown in Fig.~\ref{fig_SS}(b). This is exactly the Weyl arc protected by the nonzero second Chern number of WSs, which is shown in earlier studies \cite{Lian:2016tbn,Lian:2017thp}. Besides, in the above we further show the 3D Fermi hypersurface is also nontrivial when the WSs have nonzero second Chern numbers, which has to connect two WSs as shown in Fig.~\ref{fig_SS}(c) and \ref{fig_SS}(e).

We note that when the system has $\mathsf{TP}$ symmetry, the system will contain Yang monopoles instead of WSs, which are protected by non-abelian second Chern numbers as we discussed in last section. In this case, one will have Weyl arcs connecting Yang monopoles with opposite non-ablian second Chern numbers \cite{Lian:2017thp}.

\section{Conclusion}

In this paper, we studied the 2D WS degeneracies in generic 5D Weyl semimetals. In particular, we showed that their topological configuration -- most notably their linking configuration -- must satisfy non-trivial constraints Eqs.~\eqref{5D_WS_boundary}\eqref{5D_WS_same}\eqref{5D_WS_upper_lower}, which are 5D analogs to the famous Doubling Theorem in 3D Weyl semimetal (but also with non-trivial distinctions). Furthermore, the relation between WS linking and Yang monopole in 5D is established by showing a general relation Eq.~\eqref{C2_NA_and_C2_A_general} between the abelian and the non-abelian Chern numbers in topological band theory. More interestingly, when the 5D Weyl semimetal has a 4D surface, very rich topological surface states arise, including the 3D \emph{Fermi hypersurface} (parallel to the 1D Fermi arc in 3D Weyl semimetal) protected by the WSs in the bulk, and, on top of that, the 1D \emph{Weyl arc} protected by the linking of WSs in the bulk. The topological protection of the surface physics is closely related to the topological constraints in the bulk that we established. Brillouin zones with synthetic dimensions has been realized in cold atom systems \cite{Celi:2013gma}, so the rich surface state physics we derived maybe observed in such experiments.

We would like to make some final comments on the general mathematical framework behind the topology of band degeneracies. One early and deep result in this area is obtained for the stability of generalized Fermi surfaces (which include Weyl degeneracies) using K-theory \cite{Horava:2005jt} (other studies of the stability of gapless fermionic ground states based on Green's functions include e.g. \cite{Volovik:2011kg, Zubkov:2012ws}). More recently, the machinery of K-theory and homotopy theory has been extensively applied to understand the topology of band degeneracies. Most particularly, as we have mentioned in Section \ref{sect_Constraints}, a homotopy study of $(\mathsf{TP})^2=+1$ 3D nodal line systems \cite{Ahn:2018qgz, Wu:2018fgt} has led to constraints analogous to our ones in 5D Weyl semimetal. The connection between our simple Berry curvature method and the full machinery of homotopy theory might not be surprising -- while the celebrated Doubling Theorem is usually presented using a Berry curvature computation (as we did), originally the theorem was established using homotopy \cite{Nielsen:1980rz}. Aside from homotopy theory, we also note that recently a standard construction in homology theory (in which computations are much simpler compared to homotopy theory), the Mayer-Vietoris sequence, has been applied to understand Weyl degeneracies \cite{Mathai:2016gqc, Mathai:2016kgk}. At this point, this framework seems not to encompass some of the interesting features we discussed. In particular, it is unclear whether our constraints on WS linking can be detected with this method. Also, this method seems to be insensitive to the single component WS (case (a) of Fig.~\ref{fig_linking_constraint_ex}) which leads to a topologically protected 3D Fermi hypersurface; see panel (a) of Fig.~\ref{fig_SS}. (Such single WS can be realized by a simple model: Take a 3D Weyl semimetal model with momentum $k_\mu, \mu=1,2,3$, and then make the Weyl node separation depend on $k_4, k_5$ so that they annihilate for large $k_4, k_5$.) It would be interesting to understand what refinement of the Mayer-Vietoris approach is needed to capture the full topological information of the bulk degeneracies and surface states.

\acknowledgments
J.-Y.~C. thanks Tom\'{a}\u{s}~Bzdu\u{s}ek and Junyeong Ahn for explaining their closely related results in 3D nodal line systems \cite{Ahn:2018qgz, Wu:2018fgt}. J.-Y.~C. is supported by the Gordon and Betty Moore Foundation's EPiQS Initiative through Grant GBMF4302. B.~L. is supported by the Princeton Center for Theoretical Science at Princeton University. S.-C.~Z. is supported by the US Department of Energy, Office of Basic Energy Sciences under contract DE-AC02-76SF00515.

\section*{Appendix A}

In this Appendix we explain in detail what is meant by ``effectively two-band'' in the proof of the constraint \eqref{5D_WS_same}. Let's consider the $n$th and $(n+1)$th band which comprise a two-dimensional subspace of the full Hilbert space. Recall in the diagonalization of the Hamiltonian $H=UEU^\dagger$, the $N\times N$ unitary matrix $U$ has its $m$th column being the eigenvector $u_m$. Now let $U_{\{n, n+1\}}$ be the $N\times 2$ matrix whose two columns are $u_n$ and $u_{n+1}$. Let $R$ be an arbitrary $U(2)$ matrix rotating in this subspace, $\wt{U}_{\{n, n+1\}} = U_{\{n, n+1\}} R^\dagger$, $U_{\{n, n+1\}}=\wt{U}_{\{n, n+1\}} R$. We let the $U(2)$ matrix $R$ depend on the momentum $k$ such that its two columns have the same singularity as $W_{n+1/2}^i$ in the original $U_{\{n, n+1\}}$, and thus $\wt{U}_{\{n, n+1\}}$ does not have singularity on $W_{n+1/2}^i$ anymore. This way, $\wt{U}_{\{n, n+1\}}$ carries the topological information about the separation of this two-dimensional subspace from the remaining of the Hilbert space, while $R$, the ``effectively two-band'' part, carries the topological information about the separation within the two-dimensional subspace.

Let $r_n$ and $r_{n+1}$ be the two columns of $R$. Then $u_n = \wt{U}_{\{n, n+1\}} \, r_n$, $u_{n+1} = \wt{U}_{\{n, n+1\}} \, r_{n+1}$. The abelian Berry connection for $u_n$ can be written as
\begin{align}
A_n = -i u_n^\dagger \d u_n = r_n^\dagger \left( -i\d + \wt{A}_{\{n, n+1\}} \right) r_n
\label{A_n_separation}
\end{align}
where $\wt{A}_{\{n, n+1\}} = -iU_{\{n, n+1\}}^\dagger \, \d U_{\{n, n+1\}}$ is the $U(2)$ Berry connection on the two-band sub-bundle. Let's denote the above as $A_n=A_n^1 + A_n^2$. The first term $A_n^1=-i r_n^\dagger \d r_n$ is an ``effectively two-band''  $U(1)$ Berry connection. We want to show the second term $A_n^2 = r_n^\dagger \wt{A}_{\{n, n+1\}} r_n$ only has contribution to $(C_2)_n$ when $W_{n+1/2}^i$ is linked with $W_{n-1/2}$, but has no contribution regarding whether $W_{n+1/2}^i$ is linked with any component of $W_{n+1/2}$.

Consider the following geometric setup. Let $W_{n+1/2}^i=\partial \Sigma_{n+1/2}^i$ and let $\mathcal{V}$ be the 5D vicinity of $\Sigma_{n+1/2}$. We know $A_n$ is singular on 2D $W_{n \pm 1/2}$. Let $\mathcal{V}^\ast \equiv \mathcal{V} \backslash {W_{n \pm 1/2}}$ on which $A_n$ is defined. $A_n$ is not continuous over $\mathcal{V}^\ast$. Suppose $\mathcal{V}^\ast$ is covered by a set of charts which are 5D. Across the boundaries between the charts (the boundaries are 4D), $A_n$ in our expression \eqref{A_n_separation} is subjected to transition function, i.e. gauge transformation, of two kinds: the $U(1)$ gauge transformation
\begin{align}
r_n \rightarrow r_n \, e^{i\lambda}
\end{align}
characterizing how the $n$th band separates from the $(n+1)$th band, and the $U(2)$ gauge transformation
\begin{align}
U_{\{n, n+1\}} \rightarrow U_{\{n, n+1\}} \Lambda
\end{align}
characterizing how these two bands as a whole separate from the other bands. The discontinuity from the $U(1)$ gauge transformation extends from the singularity $W_{n+1/2}$ and is seen by $A_n^1$, while the discontinuity from the $U(2)$ gauge transformation extends from the singularity $W_{n-1/2}$ and is seen by $A_n^2$ (note that $\wt{A}_{\{n, n+1\}}$ sees the singularity $W_{n-1/2} \cup W_{n+3/2}$, but when projected by $r_n$, $A_n^2$ only sees the singularity $W_{n-1/2}$).

Thus, when considering the linking between $W_{n+1/2}^i$ with other components of $W_{n+1/2}$, we can limit ourselves to the ``effectively two-band'' part $A_n^1 = -i r_n^\dagger \d r_n$ and ignore $A_n^2$. Then, as mentioned in the main text, $(\d A_n^1)^2=(\d r_n^\dagger \: \d r_n)^2=0$ because $r_n$ and $r_n^\dagger$ together only have three real parameters (fixing an unimportant overall phase).

\section*{Appendix B}

In this Appendix we show \eqref{C2_NA_and_C2_A_general}. Although this is a special case of the more general Whitney sum formula, the proof to the formula is non-trivial, therefore we shall show this special case via direct computation. It suffices to show the following. Let $A \subset \{1, \dots, N\}$ be some subset of bands, which further separate into two disjoint subsets $B\cap C=\emptyset$, $B\cup C=A$. Let $U_A$ be the $N\times |A|$ matrix whose columns are $u_m, \ m\in A$, and its associated non-abelian Berry curvature is
\begin{align}
F_A = -i \, \d U_A^\dagger \left(1-U_A U_A^\dagger \right) \d U_A
\end{align}
and likewise for $B$ and $C$. We want to show
\begin{align}
\tr \left({F_A}^2 \right) = \tr \left({F_B}^2 \right) + \tr \left({F_C}^2 \right) - 2 \: \d K_{B, C}
\label{C2_NA_and_C2_NA_general}
\end{align}
where $K_{B, C}$ is a smooth 3-form except on values of momenta $k$ where the separation of $A$ into $B$ and $C$ becomes not well-defined (i.e. Weyl degeneracies between $B$ and $C$). Choosing $A=\{1, \dots, n\}, \ B=\{n\}, \ C=\{1, \dots, n-1\}$ leads to \eqref{C2_NA_and_C2_A_general}.

In summing over the $m$ indices in $(U_A)^\alpha_{\ m}$ for $m\in A$, we separate the summation into summations for $m\in B$ and $m\in C$. Thus,
\begin{widetext}
\begin{align}
\tr\left({F_A}^2\right) &= \tr\left( \d U_A \d U_A^\dagger \left( \mathbf{1} - U_A U_A^\dagger \right) \d U_A \d U_A^\dagger \left( \mathbf{1} - U_A U_A^\dagger \right) \right) \nonumber \\[.2cm]
& = \tr\left( \left(\d U_B \d U_B^\dagger + \d U_C \d U_C^\dagger \right) \left( \mathbf{1} - U_B U_B^\dagger - U_C U_C^\dagger \right) \left (\d U_B \d U_B^\dagger + \d U_C \d U_C^\dagger \right) \left( \mathbf{1} - U_B U_B^\dagger - U_C U_C^\dagger \right) \right) \nonumber \\[.2cm]
&= \tr\left({F_A}^2\right) + \tr({F_C}^2) - 2\: \tr \left( \d U_B^\dagger \d U_C \ \d U_C^\dagger \d U_B \right) \nonumber \\[.2cm]
& \ \ \ \ - 2\: \tr \left( \d U_C^\dagger \d U_B \ U_B^\dagger \d U_B \ U_B^\dagger \d U_C \right) - 2\: \tr \left( \d U_B^\dagger \d U_C \ U_C^\dagger \d U_C \ U_C^\dagger \d U_B \right) \nonumber \\[.2cm]
& \ \ \ \ + 2\: \tr \left( U_C^\dagger \d U_B \ \d U_B^\dagger \d U_B \ U_B^\dagger \d U_C \right) + 2\: \tr \left(U_B^\dagger \d U_C\ \d U_C^\dagger \d U_C \ U_C^\dagger \d U_B \right) \nonumber \\[.2cm]
& \ \ \ \ - 2\: \tr \left( U_C^\dagger \d U_B \ U_B^\dagger \d U_B \ \d U_B^\dagger \d U_C \right) - 2\: \tr \left( U_B^\dagger \d U_C \ U_C^\dagger \d U_C \ \d U_C^\dagger \d U_B\right) \nonumber \\[.2cm]
&= \tr\left({F_A}^2\right) + \tr({F_C}^2) - 2 \: \d K_{B, C}, \\[.3cm]
K_{B, C} & = \tr \left( U_B^\dagger \d U_C \ \d U_C^\dagger \d U_B \right) + \tr \left( U_C^\dagger \d U_B \ U_B^\dagger \d U_B \ U_B^\dagger \d U_C \right) + \tr \left( U_B^\dagger \d U_C \ U_C^\dagger \d U_C \ U_C^\dagger \d U_B \right)
\end{align}
\end{widetext}
where from the second line to the third line we expanded the terms and many of them cancelled because of $\d U_B^\dagger U_B = - U_B^\dagger \d U_B$, $\d U_C^\dagger U_B = - U_C^\dagger \d U_B$ (and likewise for $B\leftrightarrow C$).

It remains to show that, if over a sub-manifold $\mathcal{M} \subset BZ$ the separation of $A$ into $B$ and $C$ is well-defined, then $K_{B, C}$ is smooth over $\mathcal{M}$. Let $\mathcal{M}$ be covered by a set of charts. On each chart $U_B, U_C$ and hence $K_{B, C}$ are smooth. But discontinuity might arise at the boundary between two charts. Generally, across each chart boundary, $U_B$ and $U_C$ are subjected to a transition function, i.e. a gauge transformation in the gauge group $U(|B|) \times U(|C|)$:
\begin{align}
U_B \rightarrow U_B \Lambda_B, \ \ \ \ U_C \rightarrow U_C \Lambda_C.
\end{align}
But it is straightforward to verify $K_{B, C}$ is invariant under such gauge transformation (using $U_B^\dagger U_C = 0$). Hence $K_{B, C}$ is a smooth 3-form over $\mathcal{M}$. This completes the proof to the claim \eqref{C2_NA_and_C2_NA_general}.

Such separation of topological characterization when a vector bundle can be unambiguously separated into two sub-bundles is the theme behind the general Whitney sum formula.

\bibliography{Doubling_5D}

\end{document}